\documentclass[12pt]{article}
\usepackage{amsfonts,amssymb,amsmath}
\usepackage[dvips]{epsfig}
\usepackage[T1]{fontenc}
\usepackage[latin1]{inputenc}
\usepackage{graphicx}
\usepackage[english]{babel}
\usepackage{amsmath}
\usepackage{amssymb}
\usepackage{amsfonts}

\def\beq{\begin{equation}}
\def\eneq{\end{equation}}
\def\bea{\begin{eqnarray}}
\def\enea{\end{eqnarray}}

\begin{document}
\title{\bf{Generalized (2+1) dimensional black hole by Noether symmetry}}
\author{F. Darabi\thanks{Email: f.darabi@azaruniv.edu}, K. Atazadeh\thanks{Email: atazadeh@azaruniv.edu}, and A. Rezaei-Aghdam\thanks{Email: rezaei-a@azaruniv.edu}\\
{\small Department of Physics, Azarbaijan Shahid Madani University, Tabriz 53741-161, Iran.}}

\maketitle
\begin{abstract}
We use the Noether symmetry approach to find $f(R)$ theory of $(2+1)$ dimensional gravity and $(2+1)$ dimensional black hole solution consistent with this $f(R)$ gravity and the associated symmetry. We obtain $f({R})=D_1 R({n}/{n+1})({R}/{K})^{1/n}+D_2 R+D_3$, where the constant term $D_3$ plays no dynamical role. Then, we find general spherically symmetric solution for this $f(R)$ gravity which is potentially
capable of being as a black hole. Moreover, in the special case $D_1=0, D_2={1}$, namely $f(R)=R+D_3$, we obtain a generalized BTZ black hole which, other than common conserved charges $m$ and $J$, contains a new conserved charge $Q$. It is shown that this conserved charge corresponds to the freedom in the choice of the constant term $D_3$ and represents symmetry of the action under the transformation $R \rightarrow R'=R+D_3$ along the killing vector $\partial_{R}$. The ordinary BTZ black hole is obtained as the special case where $D_3$ is {\it fixed} to be proportional to the infinitesimal cosmological constant and consequently the symmetry is broken via $Q=0$. We study the thermodynamics of the generalized BTZ black hole and show that its entropy can be described by the Cardy-Verlinde formula.

\end{abstract}
Pacs:{04.70.Bw; 04.50.Kd; 04.70.Dy}
\maketitle

\section{Introduction}

General relativity in $(2+1)$-dimensional spacetime becomes a topological field theory with only a few nonpropagating degrees of freedom \cite{Mart}. The vacuum solution of $(2+1)$-dimensional gravity is necessarily flat when the cosmological constant is zero, and it can be shown that no black hole solutions exist \cite{Ida}. Moreover, the black hole thermodynamics, accounted by quantum states, is ill-defined in this low dimensional model, because of few degrees of freedom. However, it came as a great surprise when $(2+1)$-dimensional BTZ black hole solutions for a negative cosmological constant were shown to exist which can have an arbitrarily high entropy \cite{BTZ}. Indeed, Ba$\tilde{n}$ados, Teitelboim and Zanelli \cite{BTZ} have shown that $(2+1)$-dimensional gravity with a negative cosmological constant has a black hole solution, so called BTZ black hole. The BTZ black hole solution in $(2+1)$ dimensional spacetime is derived from a three dimensional action of gravity
\begin{equation}\label{ac1} 
I=\frac{1}{2}\int dx^{3} \sqrt{-g}\,
(R-2\Lambda)
\end{equation}
where $\Lambda=-l^{-2}$ is a negative cosmological constant characterized by a typical length $l$. The line element in the Schwarzschild coordinates is taken as
\begin{equation}\label{metric}
ds^2 =-f(r)dt^2 + f^{-1}(r)dr^{2}+r^2\left(d\phi
-\frac{J}{2r^2}dt\right)^2 
\end{equation} 
where
\begin{equation}\label{metric2}
f(r)=\left(-m+\frac{r^2}{l^2} +\frac{J^2}{4
r^2}\right).
\end{equation} 
This metric is stationary and axially symmetric having just two Killing vectors $\partial_t$ and $\partial_\phi$, corresponding to time displacement and
rotational symmetry and generically has no other symmetries.
Therefore, it is described by two parameters, mass $m$ and angular momentum (spin) $J$. BTZ black holes are asymptotically anti-de-Sitter (AdS) spacetime with no curvature singularity at the origin, and differ from Schwarzschild and Kerr solutions which are asymptotically flat spacetimes with curvature singularity at the origin. They describe a spacetime of constant negative
curvature, with outer and inner horizons, i.e. $r_{+}$ (event horizon) and $r_{-}$ (Cauchy horizon) subject to $J\neq 0$, respectively, given by  
\begin{equation}
r^{2}_{\pm}=\frac{l^2}{2}\left(m\pm\sqrt{m^2 -
\displaystyle{\frac{J^2}{l^2}} }\right). \label{horizon1} 
\end{equation} 
The two-parametric family of BTZ black holes, as AdS black holes, play a central role in AdS/CFT conjecture \cite{maldacena1} and in brane-world scenarios \cite{RS1,RS2}. The AdS/CFT correspondence has also been generalized for
BTZ black holes in higher curvature gravity \cite{Soda}.

Recently, $f(R)$ gravity as a modified theory of gravity has received considerable
attention concerning the current acceleration of the universe \cite{f(R)}. On the other hand, Noether symmetry is a physical criterion which allows one to select $f(R)$ gravity models which are compatible with this symmetry \cite{3}. This approach has also been used to obtain $f(T)$ gravity models respecting the Noether symmetry \cite{4}. On the other hand, new spherically symmetric solutions in $f(R)$ gravity have been obtained by Noether Symmetries
\cite{5}. In the present paper, we apply the Noether symmetry approach to obtain $(2+1)$ dimensional black hole solutions in $f(R)$ gravity which are consistent with the Noether symmetry. 

\section{$(2+1)$-dimensional $f(R)$ gravity with spherical symmetry} 

The action in the metric formalism for $(2+1)$ $f(R)$ gravity
takes the form
\begin{equation}\label{1}
I=\frac{1}{2}\int d^3x \sqrt{-g}f({R}).
\end{equation}
This action describes a theory of $(2+1)$ gravity where $f(R)$ is a typical function of the Ricci scalar $R$. In order to study the spherical solutions
we take the metric in the following form \cite{BTZ}
\bea\label{2}
ds^2=[-N^2(r)+r^2M^2(r)]dt^2+{N^{-2}(r)}{dr^2}
+2r^2M(r)dt d\phi+r^2d\phi^2,
\enea
where the radial functions $N(r)$ and $M(r)$ are to be determined as the degrees of freedom. The corresponding Ricci scalar is calculated as
\begin{equation}\label{3}
R=-\frac{1}{2r}(4rN'^2+4rNN''-r^3M'^2+8NN'),
\end{equation}
where $'$ denotes the derivative with respect to $r$.
In order to derive the field equations in the $f(R)$ gravity, we generalize
the degrees of freedom and define a canonical (point like) Lagrangian ${\cal L}={\cal L}(N, M, R, {N'}, {M'}, {R'})$ so that  ${\cal Q}=\{N, M, R\}$ is the configuration space and ${\cal TQ}=\{N, M, R, {N'}, {M'}, {R'}\}$ is the related tangent bundle on which ${\cal L}$ is defined \cite{3}. Now, we use the method of Lagrange multipliers to set $R$ as a constraint of the dynamics. To this end, by taking a suitable Lagrange multiplier
$\lambda$ and integrating by parts, the Lagrangian becomes canonical and the action takes on the following form

\bea\label{4}
{\cal S}=\int d^3x \sqrt{-g}[f({R})-\lambda(R+\frac{1}{2r}(4rN'^2+4rNN''
-r^3M'^2+8NN'))].
\enea
The variation of action with respect to $R$ gives $\lambda = f_R\equiv df/dR$,
so the action can be rewritten as
\bea\label{5}
{\cal S}=\int d^3x \sqrt{-g}[f({R})-f_R(R+\frac{1}{2r}(4rN'^2+4rNN''
-r^3M'^2+8NN'))].
\enea
Integrating by parts results in the following point-like Lagrangian 
\begin{equation}\label{6}
{\cal L}=r(f-R f_R)+\frac{r^3}{2}f_R M'^2-2f_R NN'+2rf_{RR}R'NN',
\end{equation}
where $f_{RR}\equiv d^2f/dR^2$. The equations of motion for $N$, $M$ and $R$ are obtained respectively as
\begin{equation}\label{7}
N(f_{RRR}R'^2+f_{RR}R'')=0,
\end{equation}
\begin{equation}\label{8}
({r^3}f_R M')'=0,
\end{equation}
\bea\label{9}
-rRf_{RR}+\frac{r^3}{2}f_{RR}M'^2-4f_{RR}NN'-2rf_{RR}N'^2
-2rf_{RR}NN''=0.
\enea

\section{Noether Symmetry}

Solutions for the dynamics given by the point-like canonical Lagrangian (\ref{6}) can be obtained by choosing cyclic variables which are related to some Noether symmetries. In general, a non-degenerate point-like canonical Lagrangian
${\cal L}$ depends on the variables $q^j(x^\mu)$ and on their derivatives $\partial_\nu q^j(x^\mu)$. 
Using the Euler-Lagrange equations and after some simple calculations we obtain
\begin{equation}\label{13}
\partial_\mu\left(\alpha^j\frac{\partial\mathcal {L}}{\partial\partial_\mu q^j}\right)=\alpha^j\frac{\partial \mathcal{L}}{\partial q^j}+\left(\partial_\mu\alpha^j\right)\frac{\partial\mathcal {L}}{\partial\partial_\mu q^j}=L_{\bf X}\mathcal{L},
\end{equation}
where $L_{\bf X}$ denotes the Lie derivative along the vector field ${\bf X}$ defined by
\begin{equation}\label{14}
{\bf X}=\alpha^j\frac{\partial}{\partial q^j}+\left(\partial_\mu\alpha^j\right)\frac{\partial}{\partial\partial_\mu q^j}, 
\end{equation}
which is the generator of symmetry for the dynamics derived by $\mathcal{L}$. 
This is a statement of \textit{Noether theorem} which asserts that
if $L_{\bf X}{\mathcal{L}}=0$, then the Lagrangian $\mathcal{L}$ is invariant along the vector field ${\bf X}$.
As a consequence, we can define the current 
\begin{equation}\label{15}
j^\mu=\alpha^j\frac{\partial\mathcal {L}}{\partial\partial_\mu q^j}\,,   \end{equation} 
which is conserved as 
\begin{equation}\label{16}
\partial_\mu j^\mu=0.   
\end{equation}
The presence of Noether symmetries allows one to reduce the dynamics and find out exact solutions as well as the analytic form of $f(R)$ \cite{3,5}.

\section{ $f(R)$ gravity consistent with Noether symmetry}

Following \cite{3}, we define the Noether symmetry in the present model by a vector field $X$ on the tangent space $T{\cal Q}=\left(M, N, R, {M'}, {N'},{R'}\right)$ of the configuration space ${\cal Q}=\left(M, N, R\right)$ 
\begin{equation}\label{17}
X=\alpha \frac{\partial}{\partial N}+\beta
\frac{\partial}{\partial M}+\gamma
\frac{\partial}{\partial R}+{\alpha'}\frac{\partial}{\partial {N'}}+{\beta'}\frac{\partial}{\partial {M'}}+{\gamma'}\frac{\partial}{\partial {R'}},
\end{equation}
such that 
\begin{equation}\label{18}
L_X {\cal L}=0.
\end{equation}
Therefore, a symmetry exists if one finds solutions of the equation $L_{\bf X}\mathcal{L}=0$ for the functions $\alpha$, $\beta$ and $\gamma$, where at least one of them is different from zero. Imposing (\ref{18}), we obtain the following system of partial differential equations

\begin{equation}\label{19}
R'N'\left(\alpha f_{RR}+f_{RR}N\frac{\partial \alpha}{\partial N}
+\gamma f_{RRR}N+ f_{RR}N\frac{\partial \gamma}{\partial R}\right)=0,
\end{equation}
\begin{equation}\label{20}
R'M'\left(2f_{RR}N\frac{\partial \alpha}{\partial M}+r^2 f_{R}\frac{\partial \beta}{\partial R}\right)=0,
\end{equation}
\begin{equation}\label{21}
N'M'\left(2 f_{RR}N\frac{\partial \gamma}{\partial M}+r^2 f_{R}\frac{\partial \beta}{\partial N}\right)=0,
\end{equation}
\begin{equation}\label{22}
N'\left(\alpha f_R+\gamma f_{RR}N+f_R\frac{\partial \alpha}{\partial N}N\right)=0,
\end{equation}
\begin{equation}\label{23}
M'^2\left(\frac{1}{2}\gamma f_{RR}+f_R \frac{\partial \beta}{\partial M}\right)=0,
\end{equation}
\begin{equation}\label{24}
N'^2\left(f_{RR} N\frac{\partial \gamma}{\partial N}\right)=0,
\end{equation}
\begin{equation}\label{25}
R'^2\left(f_{RR} N\frac{\partial \alpha}{\partial R}\right)=0,
\end{equation}
\begin{equation}\label{26}
M'\left(f_R N\frac{\partial \alpha}{\partial M}\right)=0,
\end{equation}
\begin{equation}\label{27}
R'\left(f_R N\frac{\partial \alpha}{\partial R}\right)=0,
\end{equation}
which is subject to the following constraint
\begin{equation}\label{28}
\gamma R f_{RR}=0.
\end{equation}
The case $f_{RR}=0$ leads to the common Einstein-Hilbert action which is
not of our interest. Moreover, we take $f_{RR}\neq0, f_{R}\neq0$ and also $M'\neq0, N'\neq0, R'\neq0$. So,  (\ref{28}) gives rise to
\begin{equation}\label{29}
\gamma =0.
\end{equation}
Moreover, (\ref{25}) and (\ref{26}) leads to
\begin{equation}\label{30}
\alpha=\alpha(N).
\end{equation}
Using (\ref{23}) and (\ref{29}) we obtain
\begin{equation}\label{31}
\frac{\partial \beta}{\partial M}=0 \Longrightarrow \beta=\beta(N, R).
\end{equation}
On the other hand, Eq.(\ref{21}) results in
\begin{equation}\label{32}
\frac{\partial \beta}{\partial N}=0 \Longrightarrow \beta=\beta(R).
\end{equation}
Finally, using (\ref{20}) and (\ref{30}) we have
\begin{equation}\label{33}
\beta=\beta_0=\mbox{Concst}.
\end{equation}
Imposing (\ref{19}) in (\ref{29}), we obtain the following result
\begin{equation}\label{34}
\alpha=\frac{A}{N},
\end{equation}
which solves Eq.(\ref{22}), $A$ being a constant. Using the above results
in (\ref{17}), the vector field $X$ becomes
\begin{equation}\label{35}
X=\frac{A}{N} \frac{\partial}{\partial N}+\beta_0
\frac{\partial}{\partial M}-\frac{AN'}{N^2}\frac{\partial}{\partial {N'}}.
\end{equation}
The conserved current (\ref{15}) is written as
\begin{eqnarray}\label{36}
j^r&=&\frac{A}{N}\frac{\partial{\mathcal{L}}}{\partial {N'}}+{\beta_0}\frac{\partial{\mathcal{L}}}{\partial {M'}}\nonumber\\
&=&-2A(f_R+rf_{RR}R')+\beta_0r^3 f_R M',
\end{eqnarray}
whose conservation through (\ref{16}) results in
\begin{equation}\label{36'}
-2A(f_R+rf_{RR}R')+\beta_0r^3 f_R M'=C_1,
\end{equation}
where $C_1$ is a constant. So, we have 
\begin{equation}\label{37}
-2A(f_{R}+r R' f_{RR})=C_1-\beta_0 C_2,
\end{equation} 
where according to (\ref{8}) 
\begin{equation}\label{38}
C_2={r^3}f_R M'=\mbox{Const}.
\end{equation}
The dynamical equation (\ref{7}) can be
written as 
\begin{equation}\label{39}
(R' f_{RR})'=0,
\end{equation} 
which gives rise to
\begin{equation}\label{40}
f_{R}=D_1 r+D_2,
\end{equation}
where $D_1, D_2$ are the constants of integration. Putting (\ref{40}) into (\ref{38}) results in
\bea\label{41}
M(r)=-{\frac {C_{{2}}{D_{{1}}}^{2}\ln  \left( D_{{1}}r+D_{{2}} \right) }{{D
_{{2}}}^{3}}}+{\frac {C_{{2}}{D_{{1}}}^{2}\ln  \left( r \right) }{{D_{
{2}}}^{3}}}
-\frac{1}{2}\,{\frac {C_{{2}}}{D_{{2}}{r}^{2}}}+{\frac {C_{{2}}D_{{
1}}}{{D_{{2}}}^{2}r}}.
\enea
In order to find $M$ and $R$ as functions of $r$, we follow the procedure
as is explained bellow. Using the fact that we are looking for the spherical
solutions, one may choose the following ansatz 
\begin{equation}\label{42}
R(r)=Kr^n,
\end{equation}
from which we find
\begin{equation}\label{43}
r=\left(\frac{R}{K}\right)^{1/n},
\end{equation}
where $K$ is a constant with appropriate dimension.
Putting this into (\ref{40}) leads to 
\begin{equation}\label{44}
f_{R}=D_1 \left(\frac{R}{K}\right)^{1/n}+D_2.
\end{equation}
Integration with respect to $R$ yields
\begin{equation}\label{45}
f({R})=D_1 R\left(\frac{n}{n+1}\right)\left(\frac{R}{K}\right)^{1/n}+D_2
R+D_3,
\end{equation}
where $D_3$ is a constant of integration. It is important to note that since $f(R)$ is not appeared in the field equation of $N(r)$, namely (\ref{9}), the constant term $D_3$ will not appear in the solution for $N(r)$, as a direct consequence of imposing the Noether symmetry. Therefore, the solutions $N(r)$ and $M(r)$ have symmetry in changing the value of $D_3$. 

\section{Generalized (2+1) dimensional black hole}

Using (\ref{41}), (\ref{42}) and (\ref{45}) in the equation of motion (\ref{9}) we obtain
\begin{eqnarray}\label{46}
N^2(r)&=&\frac{1}{4 D_2^4 r^2}[D_2 \left(C_2^2 (6 D1 r+D_2)+8 D_2^3 r 
(Pr+Q)\right)]\\ \nonumber
&+&\frac{1}{4 D_2^4 r^2}[2C_2^2D_1r\ln\frac{r}{D_1r+D_2}(2D_2+3D_1r)]-\frac{K r^{n+2}}{n^2+5 n+6},
\end{eqnarray}
where $P$ and $Q$ are constants of integrations. 
Now, (\ref{41}) and (\ref{46}) determine the spherical solutions
for the metric (\ref{2}) subject to a specific spherically symmetric Ricci scalar (\ref{42}). To explore the black hole solutions, the metric (\ref{2}) can be written in the following convenient form
\begin{equation}\label{47}
ds^2=-N^2(r)dt^2+N^{-2}(r)dr^2+r^2[M^2(r)dt+d\phi]^2.
\end{equation}
For given constants, $D_1, D_2, C_2, P, Q$ and given values for $n$, the shift function $N^2(r)$ may vanish and so the horizons may exist for those values of $r$ satisfying the following equation
\begin{eqnarray}\label{48}
D_2 \left(C_2^2 (6 D1 r+D_2)+8 D_2^3 r 
(Pr+Q)\right)+2C_2^2D_1r&\ln&\frac{r}{D_1r+D_2}(2D_2+3D_1r)\\ \nonumber
&-&\frac{4 D_2^4 r^2K r^{n+2}}{n^2+5 n+6}=0.
\end{eqnarray}
Therefore, we have found spherical solutions (\ref{41}) and (\ref{46}) capable of being as a black hole solution for $f(R)$ gravity (\ref{45}) subject to the Noether symmetry.

\section{Generalized BTZ black hole}

For the special case $D_1=0, D_2=1$, namely $f(R)=R+D_3$, we obtain 
\begin{equation}\label{49'}
N^2(r)=\frac{C_2^2}{4r^2}-\frac{K}{n^2+5 n+6}r^{n+2}+\frac{2Q}{r}+2P,
\end{equation}
where, as was expected before, $D_3$ is not shown up in the solution $N(r)$.
It is appealing to investigate whether one can recover the (2+1)-dimensional
BTZ black hole from $N^2(r)$ and $M(r)$. At first, it seems impossible because contrary to the BTZ solution, $D_3$ which can potentially play the role of a cosmological constant does not appear in the solution $N(r)$. However,
this conflict may be avoided if we use freedom in taking some arbitrary parameters appearing in the solution $N(r)$. One such appropriate arbitrary parameter is the constant $K$ through which the quantity $D_3$ can appear in the solution $N(r)$ by setting a relation between $K$ and $D_3$. 

To set this relation, we use the following procedure. We assume $n=0, Q=0$ and use the identifications $2P=-m$, $C_2=J$, and $K=-6l^{-2}$ such that $N^2(r)$ and $M(r)$ are identified with the well known solutions of the BTZ black hole \cite{BTZ}
\begin{equation}\label{49}
N^2(r)=-m+\frac{r^{2}}{l^{2}}+\frac{J^2}{4r^2},
\end{equation}
\begin{equation}\label{41'}
M(r)=-{\frac{J}{2r^2}},
\end{equation}
where $m$ and $J$, respectively are the mass and angular momentum of the
black hole, and $l^{-2}$ accounts for the cosmological constant $\Lambda$. Note that, according to (\ref{42}), we have $R=-6l^{-2}$ which
is in exact agreement with that of BTZ solution \cite{BTZ}. We know that the BTZ solution is obtained for $f(R)=R+2l^{-2}$, whereas here we have recovered the BTZ solution as the special case $n=0, Q=0$ of the generalized solution (\ref{49'}) obtained for $f(R)=R+D_3$. Hence, we have a freedom to match the two actions $f(R)=R+2l^{-2}$ and $f(R)=R+D_3$. In doing so, we assume $D_3=2l^{-2}$, and use $R=K=-6l^{-2}$ to obtain a desired relation $D_3=-\frac{K}{3}$, or $K=-3D_3$ through which the constant $D_3=2l^{-2}$ is appeared in the BTZ solution (\ref{49}).  

Now, let us assume $n=0, Q\neq0$. In this generalized case, the quantity $D_3$ is no longer fixed by $K$ or $R$, rather it is left as a continuous parameter of the symmetry. Therefore, $N^2(r)$ becomes independent of $D_3$ as follows
\begin{equation}\label{52}
N^2(r)=-m-\frac{R}{6}r^{2}+\frac{J^2}{4r^2}+\frac{2Q}{r},
\end{equation}
\begin{equation}\label{52'}
M(r)=-{\frac{J}{2r^2}}.
\end{equation}
The horizons of this generalized BTZ black hole are given by four real roots of the following equation
\begin{equation}\label{53}
-2R\,{r}^{4}-12\,m{r}^{2}+24Q r+3{J}^{2}=0.
\end{equation}
The location of surface of infinite red shift $r_{erg}$, is also obtained
as the solution of the following equation 
\begin{equation}\label{54}
\frac{2Q}{r}-\frac{R}{6}r^{2}-m=0.
\end{equation}

Considering the above results, one may conclude that the BTZ black hole is a solution corresponding to the action $f(R)=R+D_3$ equipped with a symmetry whose freedom in choosing the constant term $D_3$ is fixed as $D_3=2l^{-2}$. It is then reasonable that, similar to $m$ and $J$, we consider the constant
$Q$ as the conserved charge corresponding to the symmetry of solutions under the infinitesimal displacement of $R$ by $D_3$, as is appeared in the action $R+D_3$. In other words, it seems we are dealing with a symmetry of the action
under the transformation $R \rightarrow R+D_3$. Then, we may interpret $Q=0$ ( BTZ solutions (\ref{49}), (\ref{41'}) ) as an indication for the broken symmetry caused by fixing $D_3$.

Now, we show that such a symmetry does really exist. Although according to (\ref{17}), $\gamma=0$ indicates that no symmetry of $\partial/\partial R$ exist explicitly, however we can show that the killing vector $X$ is indeed proportional to $\partial/\partial R$. To this end, we first use (\ref{52}) and (\ref{52'}) to insert for $N(r)$, $M(r)$, and $N(r)'$ in (\ref{35}) and express all partial derivatives with respect to $r$. Then, we use (\ref{43}) to replace $r$ by $R$ and $\partial/\partial r$ by $\partial/\partial R$, respectively, so that (\ref{35}) casts in the following form 
\begin{equation}\label{35'}
X\sim \frac{\partial}{\partial R},
\end{equation}
This shows that $X$ is a killing vector field along which we have a symmetry under an infinitesimal transformation $R \rightarrow R+D_3$. 

\section{Black hole Thermodynamics}

In this section, we consider the metric (\ref{2}) with the functions given
by (\ref{52}) and (\ref{52'}) where $m>0$. 

\subsection{Thermodynamical quantities}

The mass, angular momentum  and area of the black hole are given respectively by \begin{equation}\label{58'}
m=\frac{J^2}{4r_+^2}+\frac{2Q}{r_+}-\frac{R}{6}{r_+^2}, 
\end{equation} 
\begin{equation}\label{59} 
J=2r_+\sqrt{m+\frac{R}{6}{r_+^2}-\frac{2Q}{r_+}}, 
\end{equation}
\begin{equation}\label{64} 
A_H=\frac{r_+}{2}, 
\end{equation} 
where suitable units has been used.
By employing the well-known Bekenstein-Hawking area formula \cite{BH}, the entropy of black hole is given by \begin{equation}\label{58} 
S=r_+. 
\end{equation}
We can express the mass $m$ in terms of $S, J, R$, and $Q$ as 
\begin{equation}\label{60} 
m=\frac{J^2}{4S^2}+\frac{2Q}{S}-\frac{R}{6}{S^2}. 
\end{equation}
The Hawking temperature, angular velocity and heat capacity of the black hole are given respectively by
\begin{equation}\label{61} 
T_{H}=\left[\frac{\partial m}{\partial S}\right]_{J,Q,R}=-\frac{R}{3}S-\frac{J^2}{2S^3}-\frac{2Q}{S^2}\,,
\end{equation} 
\begin{equation}\label{62} 
\Omega=\left[\frac{\partial m}{\partial J}\right]_{S,Q,R}=\frac{J}{2S^2}\,,
\end{equation} 
\begin{equation}\label{63}
C=T_H\left[\frac{\partial T_H}{\partial S}\right]^{-1}_{J,Q,R}=T_H\left(-\frac{R}{3}+\frac{3J^2}{2S^4}+\frac{4Q}{S^3}\right)^{-1}.
\end{equation} 
The thermodynamic potential conjugate to $Q$ is also obtained as
\begin{equation}\label{64'} 
\Phi_{c}=\left[\frac{\partial m}{\partial Q}\right]_{S,J}=\frac{2}{S}=A_H^{-1}\,.
\end{equation}

\subsection{Cardy-Verlinde Formula}

Verlinde has proposed a generalization of the Cardy formula from (1 + 1) dimensional conformal field theory (CFT) to $(n + 1)$-dimensional one \cite{Verlin}.
The Cardy-Verlinde formula is given by 
\begin{equation}\label{65} 
S_{CFT}=\frac{2\pi R_0}{\sqrt{ab}}\sqrt{E_C(2E-E_C)}, 
\end{equation} 
where $E$ is the total energy, $E_C$ is the Casimir energy, $R_0$ is the radius of the system, and $a$ and $b$ are arbitrary positive coefficients independent of $R_0$ and $S$. The Casimir energy is defined by the violation of Euler relation \begin{equation}\label{66} 
E_C=n(E+PV-T_H S-\Phi_{c} Q-\Omega J), 
\end{equation} 
where the pressure of the CFT is defined as $P=E/nV$. The total energy is given by the following sum 
\begin{equation}\label{67} 
E=E_E+\frac{1}{2}E_C, 
\end{equation}
where $E_E$ is the purely extensive part of the total energy $E$.
Also, the Casimir energy $E_C$ and the purely extensive part of
energy $E_E$ expressed in terms of the $R_0$ and $S$ are given by
\begin{equation}\label{68}
E_C=\frac{b}{2\pi R_0}S^{1-1/n} 
\end{equation} 
\begin{equation}\label{69}
E_E= \frac{a}{4\pi R_0}S^{1+1/n}.
\end{equation}
Using Witten's work on AdS/CFT correspondence \cite{Witten}, the Cardy-Verlinde formula (\ref{65}) can be derived by use of the thermodynamics of various black holes
with AdS asymptotic, in arbitrary dimension \cite{Ads/CFT}.

\subsection{Entropy of the generalized BTZ black hole by Cardy-Verlinde formula}

The entropy of generalized BTZ black hole with AdS asymptotic described by (\ref{52}) and (\ref{52'}) can be derived by the Cardy-Verlinde formula (\ref{65}). We obtain the Casimir energy $E_C$ using (\ref{66}) where $n=1$. In so doing, we evaluate the following terms
\begin{equation}\label{70}
T_H S=-\frac{R}{3}S^2-\frac{J^2}{2S^2}-\frac{2Q}{S},
\end{equation}
\begin{equation}\label{71}
\Phi_{c} Q=\frac{2Q}{S},
\end{equation}
\begin{equation}\label{72}
\Omega J=\frac{J^2}{2S^2}.
\end{equation}
Since the generalized black hole is asymptotically anti-de-Sitter, the
total energy is $E = m$ and the Casimir energy is obtained
\begin{equation}\label{73} 
E_C=2\left(\frac{J^2}{4S^2}+\frac{2Q}{S}\right).
\end{equation} 
On the other hand, putting $n=1$ in (\ref{68}) leads to
\begin{equation}\label{74}
E_C=\frac{b}{2\pi R_0}. 
\end{equation} 
By equating the right hand sides of (\ref{73}) and (\ref{74}), the radius $R_0$ is obtained as
\begin{equation}\label{75}
R_0=\frac{b}{4\pi}\left(\frac{J^2}{4S^2}+\frac{2Q}{S}\right)^{-1}. 
\end{equation}
Moreover, by using $PV=E$, (\ref{70}), (\ref{71}), and (\ref{72}) in (\ref{66}), the quantity $(2E-E_C)$ is evaluated as 
\begin{equation}\label{76}
2E-E_C=-\frac{R}{3}S^2.
\end{equation}
The purely extensive part of the total energy $E_E$ is then obtained by substitution
of (\ref{76}) in (\ref{67})
\begin{equation}\label{77}
E_E=-\frac{R}{6}S^2.
\end{equation}
On the other hand, putting $n=1$ in (\ref{69}) gives
\begin{equation}\label{78}
E_E= \frac{a}{4\pi R_0}S^{2}.
\end{equation}
By equating the right hand sides of (\ref{77}) and (\ref{78}), the radius $R_0$ is obtained again as
\begin{equation}\label{79}
R_0=-\frac{3a}{2\pi R}. 
\end{equation}
By using (\ref{75}) and (\ref{79}), the radius expressed in terms of the arbitrary positive coefficients $a$ and $b$ is obtained
\begin{equation}\label{80}
R_0=\frac{1}{\pi}{\sqrt{-\frac{3ab}{8R}}}\left(\frac{J^2}{4S^2}+\frac{2Q}{S}\right)^{-1/2}. \end{equation}
Substitution of (\ref{73}), (\ref{76}) and (\ref{80}) in the Cardy-Verlinde
formula (\ref{65}) gives the following result 
\begin{equation}\label{81} 
S_{CFT}=S, 
\end{equation} 
which asserts that the entropy of the generalized BTZ black
hole can be expressed in the form of Cardy-Verlinde formula.

\section{Conclusions}

In the present paper, we have obtained generalized $(2+1)$ dimensional spherical
symmetric solution in $f(R)$ gravity by using the Noether symmetry. This
solution has capability of being $(2+1)$ dimensional black hole. In a special
case, this solution casts in the form of generalized $(2+1)$ dimensional BTZ black hole. This black hole has three conserved charges as mass $m$, angular momentum $J$ and a new conserved charge $Q$ corresponding respectively to the invariance of the solution under time translation, rotation, and continuous displacement of the Ricci scalar in the action. In the same way that the killing vectors $\partial_t$ and $\partial_\phi$ with continuous space-time parameters $t$ and $\phi$ cause continuous symmetry under $t\rightarrow t'=t+\epsilon$ and $\phi \rightarrow \phi'=\phi+\delta$ ($\epsilon$ and $\delta$ being infinitesimal constants) and yield the conserved charges $m$ and $J$ respectively, the appearance of conserved charge $Q$ in this black hole is a necessary consequence of a killing vector $\partial_{R}$ (considering $R$ as a continuous quantity) which causes continuous symmetry under $R \rightarrow R'=R+D_3$ ($D_3$ being infinitesimal constant)\footnote{This may shed light on the cosmological constant problem in that why the cosmological constant is so extremely small.}. 

We have shown that the ordinary anti-de Sitter BTZ black hole within Eisntein-Hilbert theory of gravity with a negative cosmological constant is the special case $Q=0$ of this generalized BTZ black hole, where the continuous symmetry along
the vector field  $\partial_{R}$ is broken by {\it fixing} the constant term $D_3$ to be an infinitesimal cosmological constant $2l^{-2}$, for ever. In other words, it seems that what we know as the ordinary BTZ black hole, is noting but the reduction of generalized BTZ black
hole to the {\it fixed} constant hypersurface $R'=R+2l^{-2}=-4l^{-2}$.

The present study may also have important impact on the AdS/CFT correspondence
from the black hole and its thermodynamics point of view. Witten has argued that the thermodynamics of a certain conformal field theory can be identified with the thermodynamics of black holes in anti-de Sitter space \cite{Witten}. Here, we have obtained a new class of anti-de Sitter BTZ black holes in modified $f(R)$ theory of gravity subject to the Noether symmetry which has a conserved charge playing the role of a {\it geometric mass}. Hence, one may think that  the thermodynamics of a certain conformal field theory can be identified with the thermodynamics of a black hole solution obtained in the modified $f(R)$ theory of gravity consistent with such a Noether symmetry. This is an evidence for the validity of Ads/CFT correspondence in the $f(R)$ theory of gravity subject to the Noether symmetry. The rigorous study of such Ads/CFT correspondence is very appealing and needs another work.

\end{document}